\documentclass[12pt]{article}

\usepackage{graphics,epsfig}

\def\beq   {\begin{equation}}
\def\eeq   {\end{equation}}
\def\beqd  {\begin{displaymath}}
\def\eeqd  {\end{displaymath}}
\def\beqaa {\begin{eqnarray}}
\def\eeqaa {\end{eqnarray}}

\def\noi {\noindent}

\def\ti  {\tilde}

\def\sq  {\ti q}
\def\st  {\ti t}
\def\sb  {\ti b}
\def\sg  {\ti g}

\def\nt  {\tilde\chi^0}
\def\ch  {\tilde\chi^\pm}
\def\chp {\tilde\chi^+}
\def\chm {\tilde\chi^-}

\def\a   {\alpha}
\def\b   {\beta}
\def\t   {\theta}
\def\tst {\theta_{\st}}
\def\tsb {\theta_{\sb}}

\def\gev {\rm ~GeV}

\def\sz{\ifmmode{\tilde{\chi}^0} \else{$\tilde{\chi}^0$} \fi}
\def\sw{\ifmmode{\tilde{\chi}} \else{$\tilde{\chi}$} \fi}

\newcommand{\gsim}{\;\raisebox{-0.9ex}
           {$\textstyle\stackrel{\textstyle >}{\sim}$}\;}
\newcommand{\lsim}{\;\raisebox{-0.9ex}{$\textstyle\stackrel{\textstyle<}
           {\sim}$}\;}

\begin{document}
\pagestyle{empty}

\vspace*{-1cm} 
\begin{flushright}
  hep-ph/9804265\\
  Phys. Lett. B 435 (1998) 
\end{flushright}

\vspace*{1.4cm}

\begin{center}

{\Large {\bf
Bosonic decays of \boldmath{$\tilde t_2$} and \boldmath{$\tilde b_2$}
}}\\

\vspace{10mm}

{\large 
A.~Bartl$^a$, H.~Eberl$^b$, K.~Hidaka$^c$, S.~Kraml$^b$, 
T.~Kon$^d$,\\[2mm]
W.~Majerotto$^b$, W.~Porod$^a$, and Y.~Yamada$^e$}

\vspace{6mm}

\begin{tabular}{l}
$^a${\it Institut f\"ur Theoretische Physik, Universit\"at Wien, A-1090
Vienna, Austria}\\
$^b${\it Institut f\"ur Hochenergiephysik der \"Osterreichischen Akademie
der Wissenschaften,}\\
\hphantom{$^b$}{\it A-1050 Vienna, Austria}\\
$^c${\it Department of Physics, Tokyo Gakugei University, Koganei,
Tokyo 184--8501, Japan}\\
$^d${\it Faculty of Engineering, Seikei University, Musashino, Tokyo 180,
Japan}\\
$^e${\it Department of Physics, Tohoku University, Sendai 980--8578, Japan}
\end{tabular}

\end{center}

\vfill

\begin{abstract} 
We perform a detailed study of the decays of the heavier top and bottom 
squarks ($\st_2$ and $\sb_2$) in the Minimal Supersymmetric 
Standard Model (MSSM). 
We show that the decays into Higgs or gauge bosons, i.e. 
$\st_2 \to \st_1 + (h^0,H^0,A^0 \ \mbox{or} \ Z^0)$,  
$\st_2 \to \sb_{1,2} + (H^+ \ \mbox{or} \ W^+)$, and the analogous $\sb_2$ 
decays, can be dominant in a wide range of the model parameters 
due to the large Yukawa couplings and mixings of $\st$ and $\sb$. 
Compared to the conventional decays into fermions, such as 
$\st_2 \to t + (\nt_i \ \mbox{or} \ \sg)$ and $\st_2 \to b + \chp_j$, 
these bosonic decay modes can have significantly different decay 
structures and distributions. 
This could have an important impact on the search for $\st_2$ 
and $\sb_2$ and the determination of the MSSM parameters 
at future colliders.
\end{abstract}

\newpage
\pagestyle{plain}
\setcounter{page}{2}


The search for supersymmetric (SUSY) particles is one of the most 
important subjects at present and future collider experiments. 
Future colliders, such as the Large Hadron Collider (LHC), the upgraded 
Tevatron, $e^+e^-$ linear colliders, and $\mu^+\mu^-$ colliders will 
extend the discovery potential for SUSY particles to the TeV mass range 
and allow for a precise determination of the SUSY parameters.

\noi
Many phenomenological and experimental studies have been 
performed for squark ($\sq$) search \cite{ref1}. 
In most studies, production and decay of the squarks are studied 
assuming that they decay into fermions, i.~e.
a quark plus a neutralino ($\nt_k$), chargino ($\ch_j$), 
or gluino ($\sg$): 
\begin{equation}
  \sq_i^{} \to q^{(\prime)} + (\nt_k,\ch_j,\, \mbox{or}\, \sg)\, ,
  \label{eq:fmodes}
\end{equation}
with $i,j=1,2$ and $k=1,...,4$. 
However, in the Minimal Supersymmetric Standard Model 
(MSSM) \cite{ref2} the heavier squarks of the 3rd generation (i.e. 
stops ($\st_2$) and sbottoms ($\sb_2$)) can also decay into
bosons, i.~e. a lighter squark plus a gauge or 
Higgs boson \cite{ref3,ref4}:
\beq
  \begin{array}{lcl}
    \st_2 \to \st_1\,Z^0\,, &\hspace{3mm}& \sb_2 \to \sb_1\,Z^0\,, \\
    \st_2 \to \sb_i\:W^+\,, &\hspace{3mm}& \sb_2 \to \st_i\:W^-\,, 
  \end{array}
  \label{eq:wzmodes}
\eeq
and
\beq
  \begin{array}{lcl}
    \st_2 \to \st_1\:(h^0,\,H^0,\,A^0)\,, &\hspace{4mm}& 
    \sb_2 \to \sb_1\:(h^0,\,H^0,\,A^0)\,, \\
    \st_2 \to \sb_i\:H^+\,, &\hspace{4mm}& \sb_2 \to \st_i\:H^-\,.
  \end{array}
  \label{eq:hxmodes}
\eeq
Here $\sq_1^{}$ ($\sq_2^{}$) is the lighter (heavier) squark mass 
eigenstate. 
The $\st_2$ decays into $\st_1$ plus a neutral boson 
in Eqs.~(\ref{eq:wzmodes}) and (\ref{eq:hxmodes}) are possible in case 
the difference between $m_{\st_L}$ and $m_{\st_R}$ and/or the 
$\st_L$--$\st_R$ mixing are 
large enough to make the necessary mass splitting between 
$\st_1$ and $\st_2$. 
The same holds for the decays
$\sb_2\to\sb_1 + (Z^0,h^0,H^0,A^0)$.

\noi
In the present article we make a more general analysis 
than \cite{ref3,ref4}. 
We point out that the $\sq_2^{}$ decays into gauge or Higgs 
bosons of Eqs.~(\ref{eq:wzmodes}) and (\ref{eq:hxmodes})
can be dominant in a large region of the MSSM parameter space due to 
large top and bottom Yukawa couplings and 
large $\tilde{t}$ and $\tilde{b}$ mixing parameters.  
This dominance of the Higgs/gauge boson modes over the conventional 
fermionic modes of Eq.~(\ref{eq:fmodes})
could have a crucial impact on searches for 
$\st_2$ and $\sb_2$ at future colliders.


First we summarize the MSSM parameters in our analysis. 
In the MSSM the squark sector is specified by the squark mass matrix in 
the basis $(\sq_L^{},\sq_R^{})$ with $\sq=\st$ or $\sb$~\cite{ref5,ref6}
\begin{equation}
  {\cal M}^2_{\sq}= 
     \left( \begin{array}{cc} 
                m_{\sq_L}^2 & a_q m_q \\
                a_q m_q     & m_{\sq_R}^2
      \end{array} \right)       
  \label{eq:f}
\end{equation}
with
\begin{eqnarray}
  m_{\sq_L}^2 &=& M_{\ti Q}^2 
                  + m_Z^2\cos 2\beta\,(I_{3L}^q - e_q\sin^2\t_W) 
                  + m_q^2, \label{eq:g} \\
  m_{\sq_R}^2 &=& M_{\{\ti U,\ti D\}}^2  
                  + m_Z^2 \cos 2\b\, e_q\, \sin^2\t_W + m_q^2, 
                                              \label{eq:h}\\[2mm]
  a_q m_q     &=& \left\{ \begin{array}{l}
                     (A_t - \mu\cot\beta)\;m_t~~(\sq=\st)\\
                     (A_b - \mu\tan\beta)\,m_b~~(\sq=\sb) \, .
                  \end{array} \right. \label{eq:i}
\end{eqnarray}
Here $I_3^q$ is the third component of the weak isospin and $e_q$ the 
electric charge of the quark $q$.
$M_{\ti Q,\ti U,\ti D}$ and $A_{t,b}$ are soft SUSY--breaking 
parameters, $\mu$ is the higgsino mass parameter, 
and $\tan\b = v_2/v_1$ with $v_1$ $(v_2)$ being the vacuum 
expectation value of the Higgs $H_1^0$ $(H_2^0)$. 
From renormalization group equations 
we expect that $M_{\ti Q}$, $M_{\ti U}$, and $M_{\ti D}$ of the 
third generation are different from those of the first two generations 
and from each other.  
Diagonalizing the matrix (\ref{eq:f}) one gets the mass eigenstates 
$\sq_1^{}=\sq_L^{}\cos\t_{\sq}+\sq_R^{}\sin\t_{\sq}$, 
$\sq_2^{}=-\sq_L^{}\sin\t_{\sq}+\sq_R^{}\cos\t_{\sq}$ 
with the masses $m_{\sq_1}$, $m_{\sq_2}$ ($m_{\sq_1}<m_{\sq_2}$) 
and the mixing angle $\t_{\sq}$. 
As can be seen, sizable mixing effects can be expected in the stop sector 
due to the large top quark mass. Likewise, 
$\sb_L$--$\sb_R$ mixing may be important for large $\tan\beta$. \\
The properties of the charginos $\ch_i$ ($i=1,2$; $m_{\ch_1}<m_{\ch_2}$) 
and neutralinos $\nt_k$ ($k=1,...,4$; $m_{\nt_1}< ...< m_{\nt_4}$)  
are determined by the parameters $M$, $M'$, $\mu$ and $\tan\b$, 
where $M$ and $M'$ are the SU(2) and U(1) gaugino masses, respectively. 
Assuming gaugino mass unification we take $M'=(5/3)\tan^2\t_W M$ and
$m_{\sg}=(\alpha_s(m_{\sg})/\alpha_2)M$ with $m_{\sg}$ being the gluino 
mass. 
The masses and couplings of the Higgs bosons $h^0, ~H^0, ~A^0$ and $H^{\pm}$, 
including leading Yukawa corrections, are fixed by 
$m_A,~\tan\beta,~\mu,~m_t,~m_b,~M_{\ti Q},~M_{\ti U},~M_{\ti D}, 
~A_t,$ and $A_b$. 
$H^0$ ($h^0$) and $A^0$ are the heavier (lighter) CP--even and CP--odd 
neutral Higgs bosons, respectively. 
For the Yukawa corrections to the $h^0$ and $H^0$ masses 
and their mixing angle $\alpha$ we use the formulae of 
Ref.~\cite{ref9}. 
For $H^\pm$ we take $m_{H^\pm}^2 = m_A^2 + m_W^2$.

\noi
The widths of the squark decays into Higgs and gauge bosons are given by 
($i,j=1,2$; $k=1...4$) \cite{ref3}:
\beq
  \Gamma (\sq_i^{} \to \sq_j^{(\prime)}\,H_{\!k}^{}) = 
    \frac{\kappa_{ijk}^{}}{16\pi\,m_{\sq_i}^3}\;(G_{ijk})^2 ,
  \hspace{4mm}
  \Gamma (\sq_i^{} \to \sq_j^{(\prime)}\,V) = 
    \frac{\kappa_{ijV}^3}{16\pi\,m_V^2\,m_{\sq_i}^3}\;(c_{ijV})^2.
\label{eq:widths}
\eeq
Here $H_{\!k}^{}=\{h^0\!,\,H^0\!,\,A^0\!,\,H^\pm\}$ and $V=\{Z^0,W^\pm\}$. 
The $G_{ijk}$ denote the squark couplings to Higgs bosons 
and $c_{ijV}$ those to gauge bosons.
$\kappa_{ijX}^{}\equiv\kappa(m_{\sq_i}^2,m_{\sq_j^{(\prime)}}^2,m_X^2)$
is the usual kinematic factor, 
$\kappa(x,y,z)=(x^2+y^2+z^2-2xy-2xz-2yz)^{1/2}$. 
(Notice that $\Gamma (\sq_i^{} \to \sq_j^{(\prime)}\,H_{\!k}^{})$ is 
proportional to $\kappa$ whereas 
$\Gamma (\sq_i^{} \to \sq_j^{(\prime)}\,V)$ is proportional to $\kappa^3$.) 
The complete expressions for $G_{ijk}$ and $c_{ijV}$, as well as the widths 
of the fermionic modes, are given in \cite{ref3,ref10}. 
The leading terms of the squark couplings to Higgs and vector bosons 
are given in Table~1. The Yukawa couplings $h_{t,b}$ are given as 
\beq
  h_t = g\,m_t/(\sqrt{2}\,m_W\sin\b), \hspace{6mm}
  h_b = g\,m_b/(\sqrt{2}\,m_W\cos\b) \, .
\eeq
%
\noi 
\begin{table}[ht] \hrule \vspace{1mm}
\beqd
  \begin{array}{lll|lll}
    \st_1\st_2Z^0 & \!\!\sim\!\! & g\,\sin 2\tst & ~~
    \sb_1\sb_2Z^0 & \!\!\sim\!\! & g\,\sin 2\tsb 
    \\
    \st_1\st_2h^0 & \!\!\sim\!\!
                  & h_t\,(\mu\sin\a + A_t\cos\a) \cos 2\tst ~~ & ~~
    \sb_1\sb_2h^0 & \!\!\sim\!\!
                  & h_b\,(\mu\cos\a + A_b\sin\a) \cos 2\tsb 
    \\
    \st_1\st_2H^0 & \!\!\sim\!\!
                  & h_t\,(\mu\cos\a - A_t\sin\a) \cos 2\tst ~~ & ~~
    \sb_1\sb_2H^0 & \!\!\sim\!\!
                  & h_b\,(\mu\sin\a - A_b\cos\a) \cos 2\tsb 
    \\
    \st_1\st_2A^0 & \!\!\sim\!\!
                  & h_t\,(\mu\sin\b + A_t\cos\b) &  ~~
    \sb_1\sb_2A^0 & \!\!\sim\!\!
                  & h_b\,(\mu\cos\b + A_b\sin\b)  
  \end{array}
\eeqd
\beqd
  \begin{array}{llr}
     \st_1\sb_1H^\pm &\!\!\sim\!\!
                     & h_t\,(\mu\sin\b + A_t\cos\b)\sin\tst\cos\tsb 
                      +h_b\,(\mu\cos\b + A_b\sin\b)\cos\tst\sin\tsb \\
     \st_1\sb_2H^\pm &\!\!\sim\!\!
                     & -h_t\,(\mu\sin\b + A_t\cos\b)\sin\tst\sin\tsb 
                       +h_b\,(\mu\cos\b + A_b\sin\b)\cos\tst\cos\tsb \\
     \st_2\sb_1H^\pm &\!\!\sim\!\!
                     & h_t\,(\mu\sin\b + A_t\cos\b)\cos\tst\cos\tsb 
                      -h_b\,(\mu\cos\b + A_b\sin\b)\sin\tst\sin\tsb \\
     \st_2\sb_2H^\pm &\!\!\sim\!\!
                     & -h_t\,(\mu\sin\b + A_t\cos\b)\cos\tst\sin\tsb 
                       -h_b\,(\mu\cos\b + A_b\sin\b)\sin\tst\cos\tsb 
  \end{array}   
\eeqd
\beqd
   \st_i\sb_jW^\pm \sim g \,\left(\! \begin{array}{rr} 
       \cos\tst \cos\tsb & -\cos\tst \sin\tsb \\
      -\sin\tst \cos\tsb &  \sin\tst \sin\tsb
   \end{array}\right)_{ij}
\eeqd
\hrule
\caption{\em Squark couplings to Higgs and vector bosons (leading terms).}
\label{table1}
\end{table} 
As can be seen, the $\sq_1^{}\sq_2^{}Z^0$ couplings take their maximum 
for full $\sq_L^{}$--$\sq_R^{}$ mixing ($\t_{\sq}\to \pi/4$ or $3\pi/4$) 
and vanish in case of no mixing.  
The reason is that the $Z^0$ couples only to 
$\sq_L^{\dagger}\sq_L^{}$ and $\sq_R^{\dagger}\sq_R^{}$.
On the other hand, the $W^\pm$ couples only to the left components of 
the squarks. 
In contrast to that, Higgs bosons couple mainly to 
$\sq_L^{}$--$\sq_R^{( ')}$ combinations. 
These couplings are proportional to the Yukawa couplings 
$h_{t,b}$ and the parameters $A_{t,b}$ and $\mu$, 
as can be seen in Table~1.  
Notice here that the $\sq_1^{}\sq_2^{}h^0$ and $\sq_1^{}\sq_2^{}H^0$ 
couplings have a factor $\cos 2\t_{\sq}$ (which decreases with increase 
of the $\sq$--mixing) while the $\sq_1^{}\sq_2^{}A^0$ couplings 
do not depend explicitly on the squark mixing angles. 
Hence the $\st_1\st_2 A^0$ coupling (and the $\sb_1\sb_2 A^0$ coupling 
for large $\tan\b$) can be especially strong in case $A_t$ ($A_b$) 
and $\mu$ are large. 
Notice also that the squark mixing angles themselves depend on $A_{t,b}$, 
$\mu$, and $\tan\b$. Moreover, $\sq_L^{}$--$\sq_R^{}$ mixing enhances the 
splitting of the $\sq$ mass eigenvalues, which in turn can have an important 
influence on the phase space of the $\sq_2^{}$ decays. 


We now turn to the numerical analysis of the $\st_2$ and $\sb_2$ decay 
branching ratios. 
For this, we calculate the widths of all possibly important 
2--body decay modes (3--body decays \cite{ref10} are negligible 
in this study):
\beqaa
  \st_2 &\to& t\sg,~ t\nt_k,~ b\chp_j,~ \st_1 Z^0,~ \sb_i W^+,~
              \st_1 h^0,~ \st_1 H^0,~ \st_1 A^0,~ \sb_i H^+,
                \nonumber \\
  \sb_2 &\to& b\sg,~ b\nt_k,~ t\chm_j,~ \sb_1 Z^0,~ \st_i W^-,~
              \sb_1 h^0,~ \sb_1 H^0,~ \sb_1 A^0,~ \st_i H^-.
                \nonumber 
\eeqaa
We take  
$m_t=175$ GeV, $m_b=5$ GeV, $m_Z^{}=91.2$ GeV, $\sin^2\t_W =0.23$, 
$m_W^{} = m_Z^{}\cos\t_W$, $\alpha(m_Z^{})=1/129$, and 
$\alpha_s(m_Z^{})=0.12$ 
[with $\alpha_s(Q)=12\pi/((33-2n_f)\ln(Q^2/\Lambda_{n_f}^2))$, 
$n_f$ being the number of quark flavors].
In order not to vary many parameters we choose
$M_{\tilde{Q}}$ = $\frac{9}{8} M_{\tilde{U}}$ = $\frac{9}{10} M_{\tilde{D}}$ 
and $A_t=A_b \equiv A$ for simplicity. 
Moreover, we fix $M=300$ GeV (i.e. $m_{\sg} = 820$~GeV) and $m_A=150$ GeV.
Thus we have as free parameters $M_{\ti Q}$, $\mu$, $\tan\b$, and $A$.
In the plots we impose the following conditions:
\renewcommand{\labelenumi}{(\roman{enumi})} 
\begin{enumerate}
  \item $m_{\ch_1} > 100$ GeV,
  \item $m_{\nt_1} > 70$ GeV,
  \item $m_{\st_1,\sb_1} > m_{\nt_1}$,
  \item $m_{h^0} > $ 80 GeV,
  \item $\Delta\rho\,(\st\!-\!\sb) < 0.0016$ \cite{ref14}
        using the formula of \cite{ref15}, and
  \item $A_t^2 < 3\,(M_{\ti Q}^2 + M_{\ti U}^2 + m_{H_2}^2)$ and 
        $A_b^2 < 3\,(M_{\ti Q}^2 + M_{\ti D}^2 + m_{H_1}^2)$ with \\
        $m_{H_2}^2=(m_A^2+m_Z^2)\cos^2\b-\frac{1}{2}\,m_Z^2$ and 
        $m_{H_1}^2=(m_A^2+m_Z^2)\sin^2\b-\frac{1}{2}\,m_Z^2$ \\
        (approximately necessary conditions to avoid colour and electric 
        charge breaking global minimum \cite{ref16}).
\end{enumerate}
Conditions (i)--(iv), along with $m_{\sg} = 820$ GeV, satisfy the 
experimental bounds on $\tilde\chi_1^+$, $\tilde\chi_1^0$,
$\st$, $\sb$, and $h^0$ from LEP2 
\cite{ref16a,ref11,ref13} 
and Tevatron \cite{ref12}.
Conditions (v) and (vi) constrain the $\st$ and $\sb$ mixings 
significantly.

\noi
In Fig.~\ref{fig:one} we plot the contour lines for the branching ratios 
of the Higgs boson modes and the gauge boson modes, 
i.~e. ${\rm BR}(\sq_2^{}\to\sq^{( \prime)}H)$ $\equiv$ 
$\sum {\rm BR}\big[\:\sq_2^{}\to \sq_1^{} + (h^0,H^0,A^0),$ 
                                $\sq'_{1,2} + H^\pm \,\big]$ 
and ${\rm BR}(\sq_2^{}\to\sq^{( \prime)}V)$ $\equiv$ 
$\sum {\rm BR}\big[\:\sq_2^{}\to \sq_1^{} + Z^0,$ 
                                $\sq'_{1,2} + W^\pm \,\big]$   
with $\sq=\st$ or $\sb$,  
in the $\mu$--$A$ plane for $M_{\tilde{Q}} = 500$~GeV and $\tan\beta = 3$. 
We see that the $\st_2$ and $\sb_2$ decays into bosons are dominant 
in a large region of the MSSM parameter space,  
in particular for large $|\mu|$ and/or $|A|$. 
Note here the dependence on the signs of $A$ and $\mu$. 
We have obtained a similar result for large $\tan\beta$.

\noi
In Fig.~\ref{fig:two} we show the individual branching ratios of 
the $\st_2$ and $\sb_2$ decays into bosons 
as a function of $\mu$ 
for $M_{\tilde{Q}} = 500$~GeV, $A = 600$~GeV, 
(a) $\tan\beta = 3$ and (b,\,c) $\tan\beta = 30$.
(We plot only branching ratios larger than 1\%.)
In case of $\tan\b=3$ (Fig.~2a) the bosonic decays of $\st_2$ 
are dominant (${\rm BR}\gsim 50\%$) for $\mu\lsim~-400\gev$ because 
(i) $(A_t-\mu\cot\b)\,m_t$, the off--diagonal element of the stop mass 
matrix, is large enough to induce the necessary mixing and mass 
splitting for the stops, and
(ii) for relatively large $|\mu|$ the decays into higgsino--like 
neutralinos ($\nt_{3,4}$) and chargino ($\ch_2$) are kinematically 
suppressed or forbidden. 
The branching ratio of the $\st_1 Z^0$ mode has its maximum 
at $\mu = -600\gev$ where $m_{\st_1}=332\gev$, $m_{\st_2}=628\gev$, 
and $\t_{\st}\simeq 130^{\circ}$. 
For further decreasing $\mu$ the decay into $\st_1 A^0$ quickly gains 
importance as the $\st_1\st_2 A^0$ coupling is 
$\sim h_t(0.3\,A_t+0.95\,\mu)$.    
The $\sb_1 W^+$ mode has a branching ratio of 10\% to 16\% for 
$\mu \lsim -400\gev$ because $\sb_1\sim\sb_L$ and $m_{\sb_1}\simeq 500\gev$.
For positive $\mu$ the bosonic $\st_2$ decays are kinematically 
suppressed or even forbidden due to insufficient stop mass splitting.  
Similar arguments hold for the $\sb_2$ decays. 
Here only the decays into $\st_1 H^-$ and $\st_1 W^-$ are 
kinematically accessible, which can be seen in Figs.~1c and 1d. \\
For $\tan\b=30$ (Figs.~2b and 2c) there is a large mixing also in the 
sbottom sector (which is proportional to $\mu\tan\b$ while 
$\t_{\st}\sim 128^{\circ}$, see Eq.~(\ref{eq:i})). 
Hence the $\st_2$ and $\sb_2$ decay branching ratios show a much 
weaker dependence on the sign of $\mu$. 
For $|\mu|\gsim 400\gev$ the decays into bosons clearly dominate. 
Notice the importance of the decays $\st_2\to\sb_1+(H^+,W^+)$ 
in Fig.~2b and  $\sb_2\to\sb_1 +(A^0,Z^0)$ in Fig.~2c. 
This is mainly due to the large mixing (and hence large mass splitting)
in the sbottom sector. 
Notice also that, in general, when the decays into $Z^0$ and 
$A^0$ are kinematically allowed, those into $h^0$ and $H^0$ are also
possible. The latter decays are, however, practically negligible in 
this example because they are suppressed by a factor $\cos^2 2\t_{\sq}$.
(Here note that ${\rm BR}(\st_2\to\st_1 h^0)$ and 
${\rm BR}(\st_2\to\st_1 H^0)$ can be about $\sim 10\%$ 
for other values of the MSSM parameters; see e.g. Fig.~4.)

\noi 
In Fig.~\ref{fig:three} we show the $\tan\beta$ dependence 
of the bosonic $\st_2$ and $\sb_2$ decay branching ratios 
for $M_{\tilde{Q}} = 500$~GeV, $A = 600$~GeV, and $\mu = - 700$~GeV. 
Here it can be seen explicitly how the $\sq_2^{}$ decays into the 
lighter sbottom $\sb_1$ plus a gauge or Higgs boson become important 
with increasing $\tan\b$.

\noi
In Fig.~\ref{fig:four} we show the $M_{\tilde{Q}}$ dependence 
of the branching ratios 
of the bosonic $\st_2$ and $\sb_2$ decays for
$A = 400$~GeV, $\mu = - 1000$~GeV, and $\tan\beta = 3$.
In this case we have 
$(m_{\sz_1}, m_{\sw_1^+}, m_{\sg})$ = (150, 302, 820) GeV. 
We see that the bosonic modes dominate the $\st_2$ and $\sb_2$ decays
in a wide range of $M_{\tilde{Q}}$. 
(Notice that the decay into a gluino is dominant above its threshold.)

\noi
For large $m_A$ the decays into $H^0$, $A^0$, and $H^\pm$ are 
phase--space suppressed. However, the remaining gauge boson modes can 
still be dominant. 
For $M_{\tilde Q}=500$ GeV, $\mu=1000$ GeV, $A=1000$ GeV, and 
$\tan\b=3$ we have, for instance, 
($m_{\st_1}$, $m_{\st_2}$, $m_{\sb_1}$, $m_{\sb_2}$, $m_{\sw_1^+}$, 
$m_{\sz_1}$) = (365, 610, 501, 558, 294, 148) GeV with 
${\rm BR}(\st_2\to\sq H)$ = (68, 58, 32, 1)\% 
and ${\rm BR}(\st_2^{}\to\sq V)$ = (18, 24, 39, 57)\% 
for $m_A = (110,\,200,\,240,\, >\!250)$ GeV.
As for the dependence on the parameter $M$,
our results do not change significantly for smaller values of $M$, 
as long as decays into gluino are kinematically forbidden.    


Let us now discuss the signatures of the $\st_2$ and $\sb_2$ decays. 
Typical signals of the decays into bosons  
(Eqs.~(\ref{eq:wzmodes}) and (\ref{eq:hxmodes})) and of 
those into fermions (Eq.~(\ref{eq:fmodes})) are shown in 
Table~1 of Ref.~\cite{ref22}. 
In principle, the final states of both types of decays can be identical.  
For example, the final state of the decay chain 
(a) $\st_2$ $\to$ $\st_1$ + $(h^0,H^0,A^0$~or~$Z^0)$ 
    $\to$ $(t \sz_1)$ + $(b \bar{b})$ 
    $\to$ $(b q \bar{q}' \sz_1)$ + $(b \bar{b})$ 
has the same event topology as that of 
(b) $\st_2$ $\to$ $t + \sz_{2,3,4}$ 
    $\to t$ + $((h^0,H^0,A^0$~or~$Z^0)$ + $\sz_1)$ 
    $\to$ $t + (b\bar{b} \sz_1)$ 
    $\to$ $(b q\bar{q}')$ + $(b\bar{b} \sz_1)$. 
Likewise 
(c) $\st_2$ $\to$ $\sb_{1,2}$ + $(H^+$~or~$W^+)$ 
    $\to$ $(b \sz_1)$ + $(q \bar{q}')$ 
has the same event topology as 
(d) $\st_2$ $\to$ $b + \sw_{1,2}^+$ 
    $\to$ $b + ((H^+$~or~$W^+)$ + $\sz_1)$ 
    $\to$ $b + (q \bar{q}' \sz_1)$. 
However, the decay structures and kinematics of the two modes 
(a) and (b) ((c) and (d)) are quite different from each other, 
since the $\sz_1$ (supposed to be the lightest supersymmetric 
particle (LSP) and hence a missing particle in case of R--parity 
conservation) is emitted from $\st_1$ and $\sz_{2,3,4}$ 
($\sb_{1,2}$ and $\sw_{1,2}^+$), respectively. 
This could result in significantly different event distributions 
(e.g. missing energy--momentum distribution) of the $\sq_2^{}$ decays 
into gauge or Higgs bosons compared to the decays into fermions.
Hence the possible dominance of the former decay modes could have 
an important impact on the search for $\st_2$ and $\sb_2$, 
and on the measurement of the MSSM parameters. 
Therefore, the effects of the bosonic decays should be included 
in the Monte Carlo studies of $\st_2$ and $\sb_2$ decays. 


In conclusion,   
we have shown that the $\tilde{t}_2$ and $\tilde{b}_2$ decays into 
Higgs or gauge bosons (such as 
$\tilde{t}_2$ $\to$ $\tilde{t}_1$ + $(h^0,H^0,A^0$~or~$Z^0)$ and 
$\tilde{t}_2$ $\to$ $\tilde{b}_{1,2}$ + $(H^+$~or~$W^+))$ 
can be dominant 
in a wide range of the MSSM parameter space due to 
large Yukawa couplings and mixings of $\tilde{t}$ and $\tilde{b}$. 
Compared to the conventional fermionic modes (such as 
$\tilde{t}_2 \to t + (\tilde{\chi}_i^0~{\rm or}~\tilde{g})$ and 
$\tilde{t}_2 \to b + \tilde{\chi}_j^+$), 
these bosonic decay modes can have significantly different 
decay structures and distributions. 
This could have an important impact 
on the searches for $\tilde{t}_2$ and $\tilde{b}_2$ and 
on the determination of the MSSM parameters at future colliders.

\section*{Acknowledgements}

The work of A.B., H.E., S.K., W.M., and W.P. was supported by 
the ``Fonds zur F\"orderung der wissenschaftlichen Forschung'' 
of Austria, project no. P10843--PHY.
The work of Y.Y. was supported in part by the Grant--in--aid for Scientific
Research from the Ministry of Education, Science, and Culture of
Japan, No.~10740106, and by Fuuju--kai Foundation.



\setlength{\unitlength}{1mm}


\noindent 
\begin{figure}
\begin{picture}(150,120)
\put(7,65){\mbox{\epsfig{figure=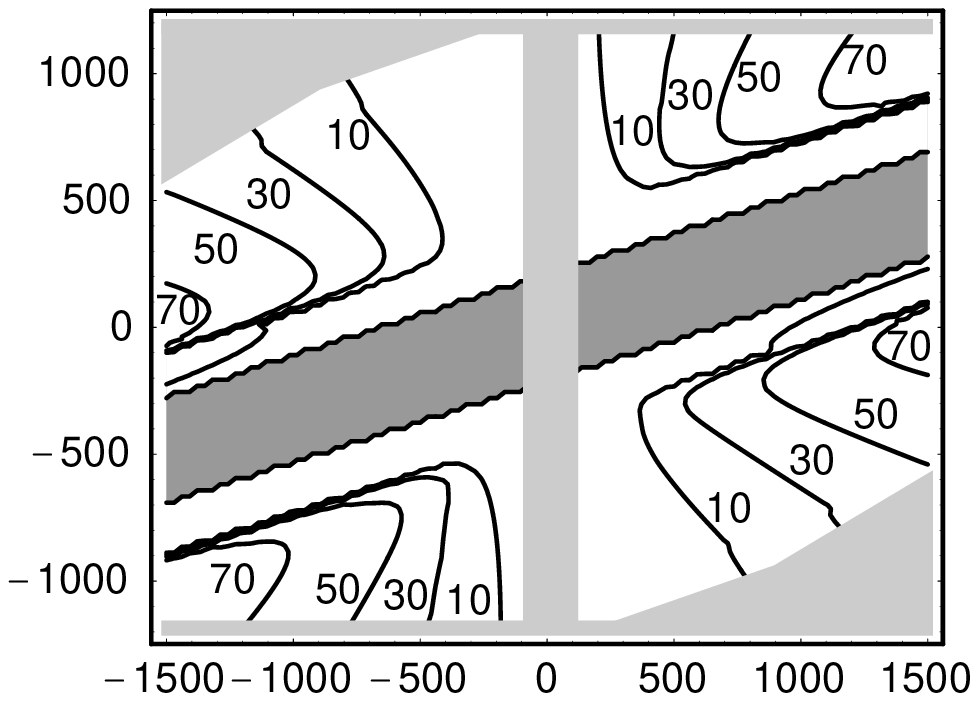,height=5.6cm}}}
\put(77,65){\mbox{\epsfig{figure=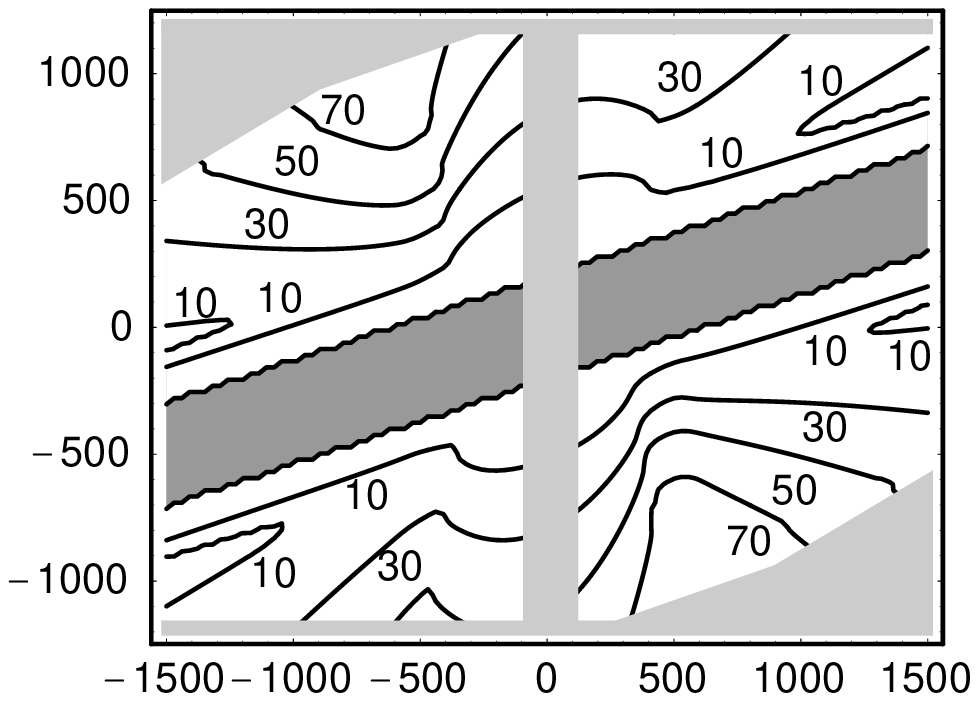,height=5.6cm}}}
\put(7,1){\mbox{\epsfig{figure=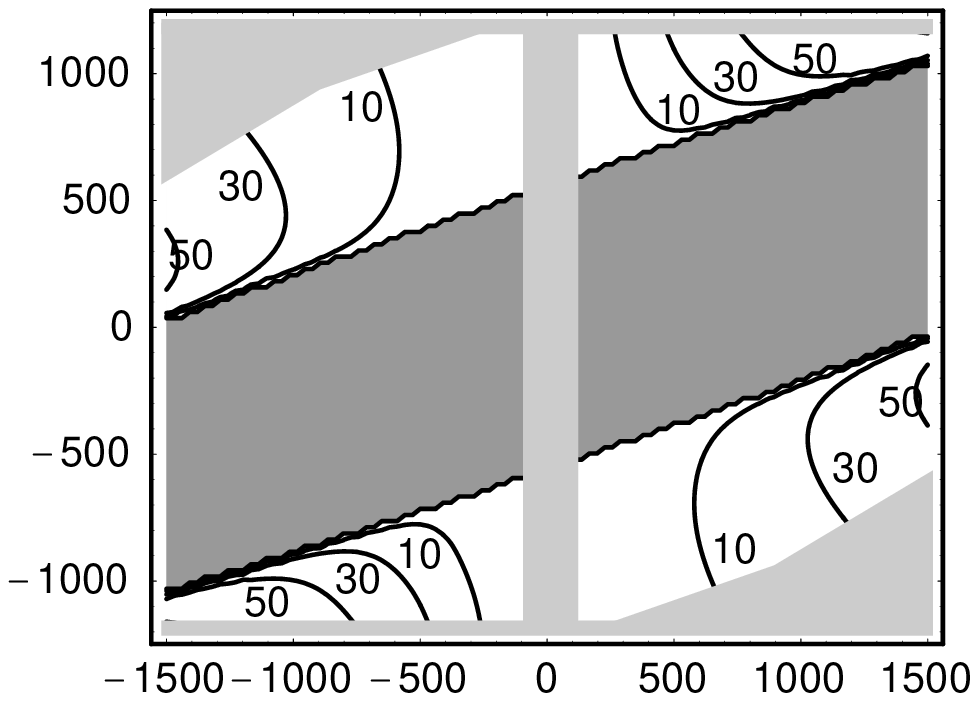,height=5.6cm}}}
\put(77,1){\mbox{\epsfig{figure=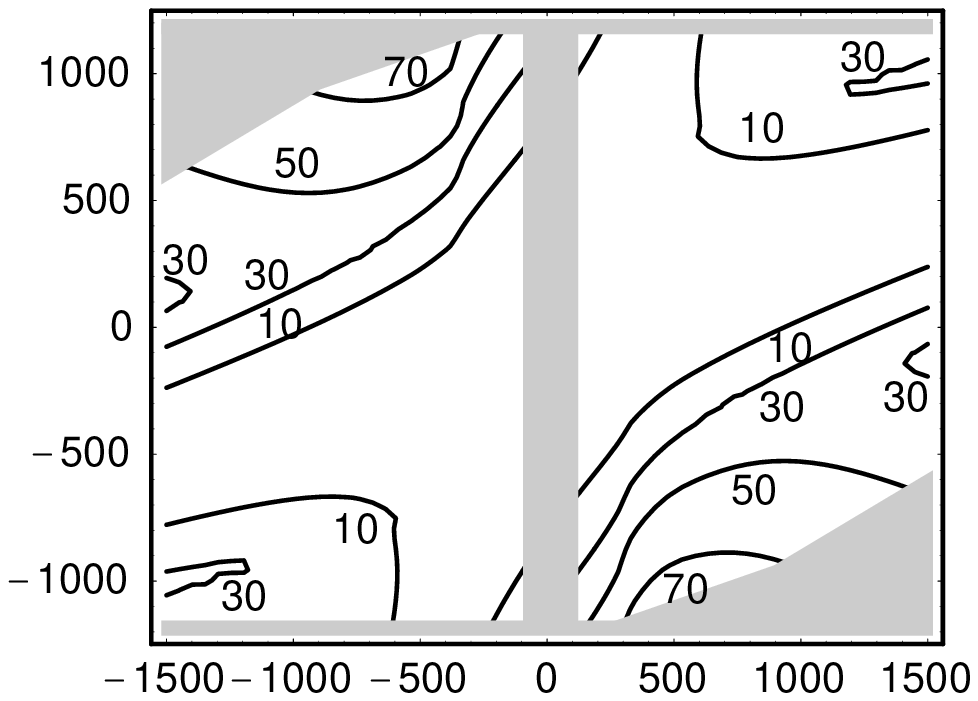,height=5.6cm}}}
\put(39,63){\mbox{$\mu~[{\rm GeV}]$}}
\put(109,63){\mbox{$\mu~[{\rm GeV}]$}}
\put(37,-1){\mbox{$\mu~[{\rm GeV}]$}}
\put(109,-1){\mbox{$\mu~[{\rm GeV}]$}}
\put(6,88){\makebox(0,0)[br]{{\rotatebox{90}{$A~[{\rm GeV}]$}}}}
\put(6,24){\makebox(0,0)[br]{{\rotatebox{90}{$A~[{\rm GeV}]$}}}}
\put(21,112){\mbox{\bf a}}
\put(91,112){\mbox{\bf b}}
\put(21,48){\mbox{\bf c}}
\put(91,48){\mbox{\bf d}}
\end{picture}
\caption{Branching ratios (in \%) of $\st_2$ and $\sb_2$ decays 
in the $\mu$--$A$ plane for $A_t=A_b\equiv A$, 
$M_{\ti Q}=500\gev$, $M_{\ti U}=444\gev$, $M_{\ti D}=556\gev$, 
$M=300\gev$, $m_A=150\gev$, and $\tan\beta=3$. 
(a) $\sum {\rm BR} 
\big[\: \st_2\to \st_1 + (h^0,H^0,A^0),~ \sb_{1,2} + H^+ \,\big]$, 
(b) $\sum {\rm BR}
\big[\: \st_2\to \st_1 + Z^0,~ \sb_{1,2} + W^+ \,\big]$, 
(c) ${\rm BR}\big[\: \sb_2\to \st_1 + H^- \,\big]$, 
(d) ${\rm BR}\big[\: \sb_2\to \st_1 + W^- \,\big]$. 
In the dark grey areas the decays are kinematically not allowed; the 
light grey areas are excluded by the conditions (i) to (vi) 
given in the text.}
\label{fig:one}
\end{figure}


\noindent 
\begin{figure}
\begin{center}
\begin{picture}(105,185)
\put(10,125){\mbox{\epsfig{figure=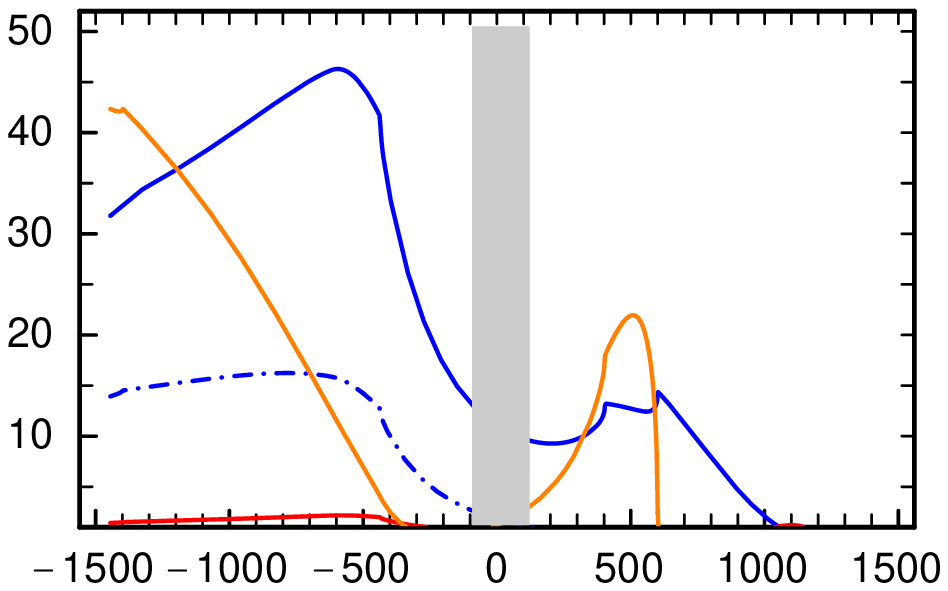,height=5cm}}}
\put(10,65){\mbox{\epsfig{figure=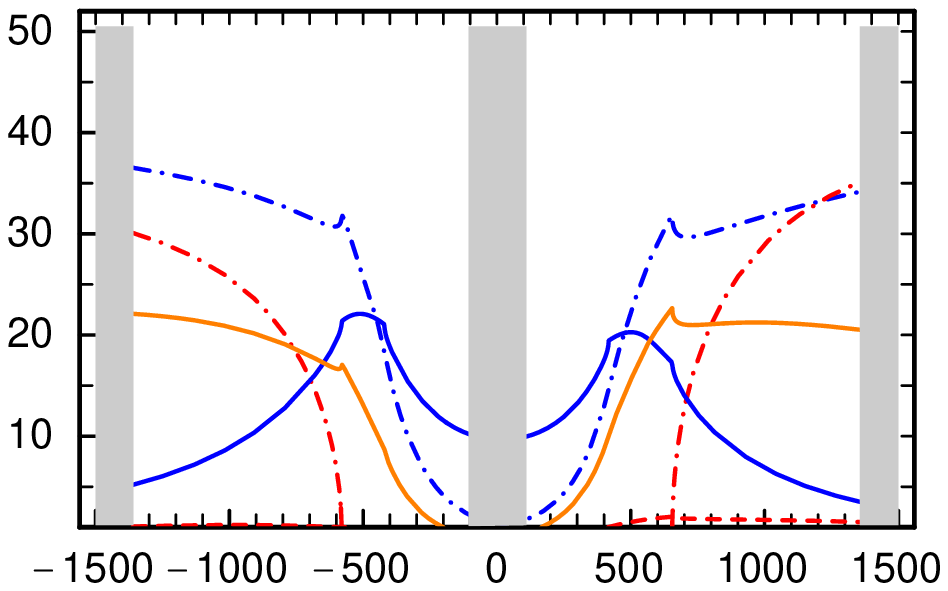,height=5cm}}}
\put(10,5){\mbox{\epsfig{figure=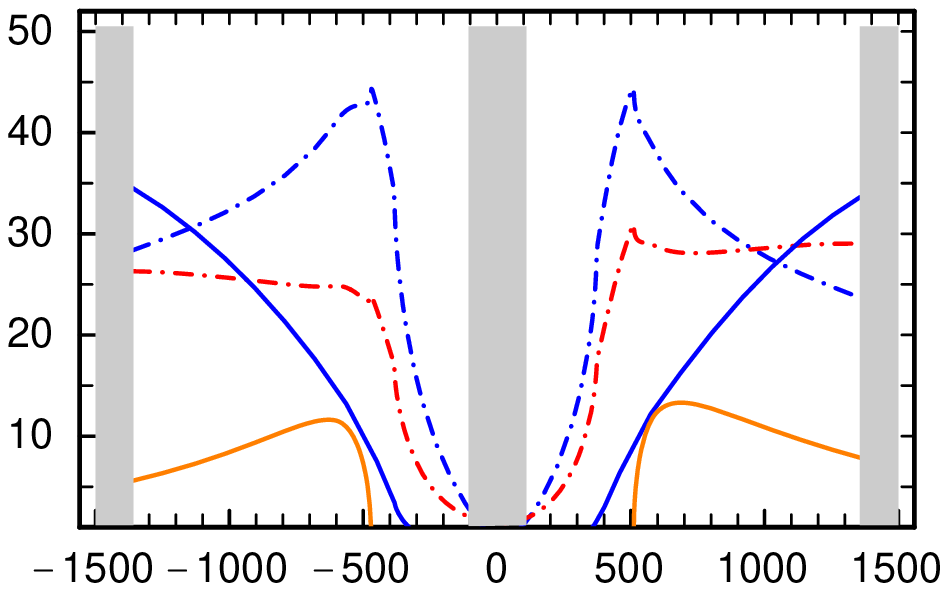,height=5cm}}}
\put(47,121){\mbox{$\mu~[{\rm GeV}]$}}
\put(47,61){\mbox{$\mu~[{\rm GeV}]$}}
\put(47,1){\mbox{$\mu~[{\rm GeV}]$}}
\put(9,142){\makebox(0,0)[br]{{\rotatebox{90}{BR\,($\st_2$)~[\% ]}}}}
\put(9,82){\makebox(0,0)[br]{{\rotatebox{90}{BR\,($\st_2$)~[\% ]}}}}
\put(9,22){\makebox(0,0)[br]{{\rotatebox{90}{BR\,($\sb_2$)~[\% ]}}}}
\put(80,169){\mbox{\bf a}}
\put(80,109){\mbox{\bf b}}
\put(80,49){\mbox{\bf c}}
\put(29,133){\mbox{\scriptsize $\st_1 h^0$}}
\put(33,154){\mbox{\scriptsize $\st_1 A^0$}}
\put(44.5,164){\mbox{\scriptsize $\st_1 Z^0$}}
\put(22,145){\mbox{\scriptsize $\sb_1 W^+$}}
\put(62,151){\mbox{\scriptsize $\st_1 A^0$}}
\put(74,136){\mbox{\scriptsize $\st_1 Z^0$}}
\put(29,102){\mbox{\scriptsize $\sb_1 W^+$}}
\put(31,93){\mbox{\scriptsize $\sb_1 H^+$}}
\put(24,85){\mbox{\scriptsize $\st_1 A^0$}}
\put(24,78){\mbox{\scriptsize $\st_1 Z^0$}}
\put(69,98.5){\mbox{\scriptsize $\sb_1 W^+$}}
\put(76,91){\mbox{\scriptsize $\sb_1 H^+$}}
\put(77,85){\mbox{\scriptsize $\st_1 A^0$}}
\put(75,78){\mbox{\scriptsize $\st_1 Z^0$}}
\put(70,72.5){\mbox{\scriptsize $\st_1 H^0$}}
\put(32,45){\mbox{\scriptsize $\st_1 W^-$}}
\put(67,45){\mbox{\scriptsize $\st_1 W^-$}}
\put(35,33){\mbox{\scriptsize $\st_1 H^-$}}
\put(65,31){\mbox{\scriptsize $\st_1 H^-$}}
\put(30,24){\mbox{\scriptsize $\sb_1 Z^0$}}
\put(73,25){\mbox{\scriptsize $\sb_1 Z^0$}}
\put(24,17.5){\mbox{\scriptsize $\sb_1 A^0$}}
\put(74,15){\mbox{\scriptsize $\sb_1 A^0$}}
\end{picture}
\end{center}
\caption{ $\mu$ dependence of $\st_2$ and $\sb_2$ decay branching ratios 
for $M_{\ti Q}=500\gev$, $M_{\ti U}=444\gev$, $M_{\ti D}=556\gev$, 
$A_t=A_b=600\gev$, $M=300\gev$, $m_A=150\gev$,  
(a) $\tan\beta=3$, (b) and (c) $\tan\beta=30$. 
The grey areas are excluded by the conditions (i) to (vi) given in the 
text.}
\label{fig:two}
\end{figure}


\noindent 
\begin{figure}
\begin{center}
\begin{picture}(105,115)
\put(10,65){\mbox{\epsfig{figure=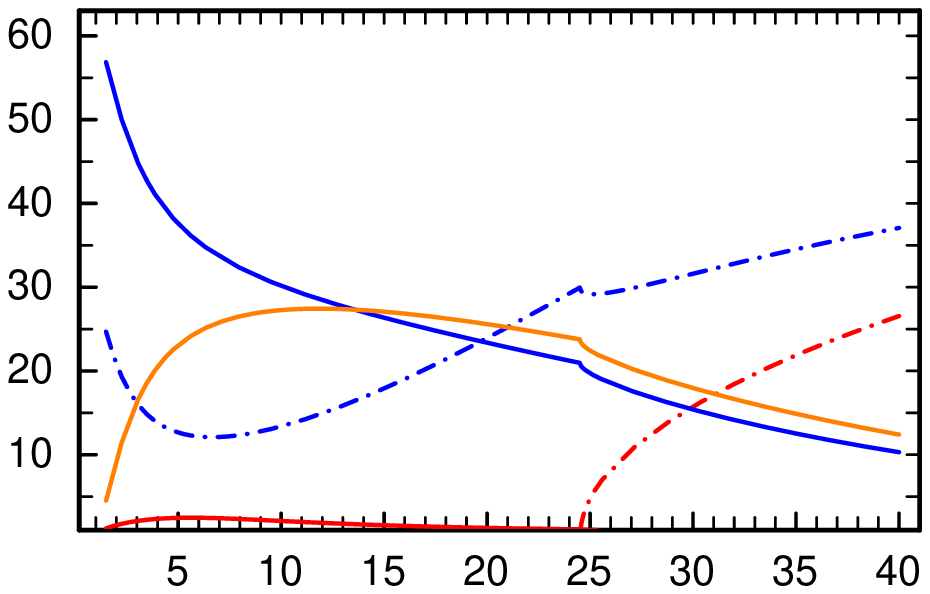,height=5cm}}}
\put(10,5){\mbox{\epsfig{figure=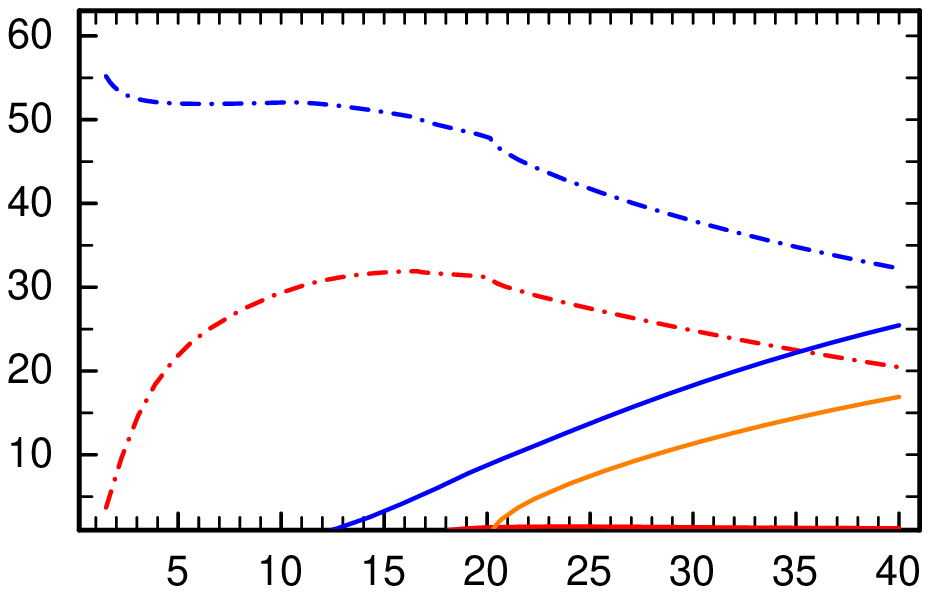,height=5cm}}}
\put(49,61){\mbox{$\tan\beta$}}
\put(49,1){\mbox{$\tan\beta$}}
\put(9,82){\makebox(0,0)[br]{{\rotatebox{90}{BR\,($\st_2$)~[\% ]}}}}
\put(9,22){\makebox(0,0)[br]{{\rotatebox{90}{BR\,($\sb_2$)~[\% ]}}}}
\put(27,100){\mbox{\scriptsize $\st_1 Z^0$}}
\put(25,89){\mbox{\scriptsize $\st_1 A^0$}}
\put(62,93){\mbox{\scriptsize $\sb_1 W^+$}}
\put(34,73){\mbox{\scriptsize $\st_1 h^0$}}
\put(76,88){\mbox{\scriptsize $\sb_1 H^+$}}
\put(63,41){\mbox{\scriptsize $\st_1 W^-$}}
\put(34,33){\mbox{\scriptsize $\st_1 H^-$}}
\put(53,20){\mbox{\scriptsize $\sb_1 Z^0$}}
\put(76,16){\mbox{\scriptsize $\sb_1 A^0$}}
\put(66,12){\mbox{\scriptsize $\sb_1 h^0$}}
\end{picture}
\end{center}
\caption{ $\tan\beta$ dependence of $\st_2$ and $\sb_2$ decay 
branching ratios for $M_{\ti Q}=500\gev$, $M_{\ti U}=444\gev$, 
$M_{\ti D}=556\gev$, $A_t=A_b=600\gev$, $\mu=-700\gev$, $M=300\gev$, 
and $m_A=150\gev$. }
\label{fig:three}
\end{figure}


\begin{figure}
\begin{center}
\begin{picture}(105,115)
\put(11.5,65){\mbox{\epsfig{figure=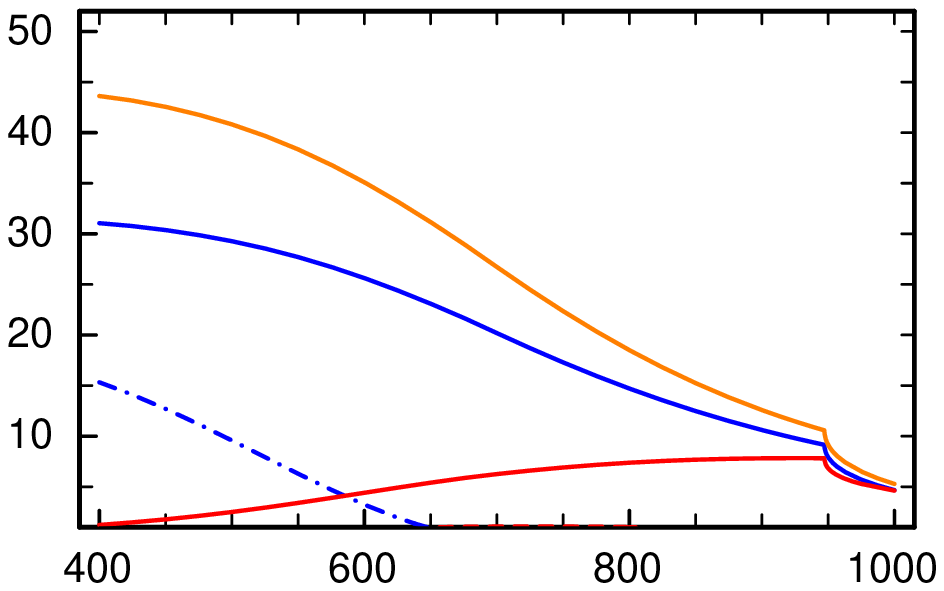,height=5cm}}}
\put(11.5,5){\mbox{\epsfig{figure=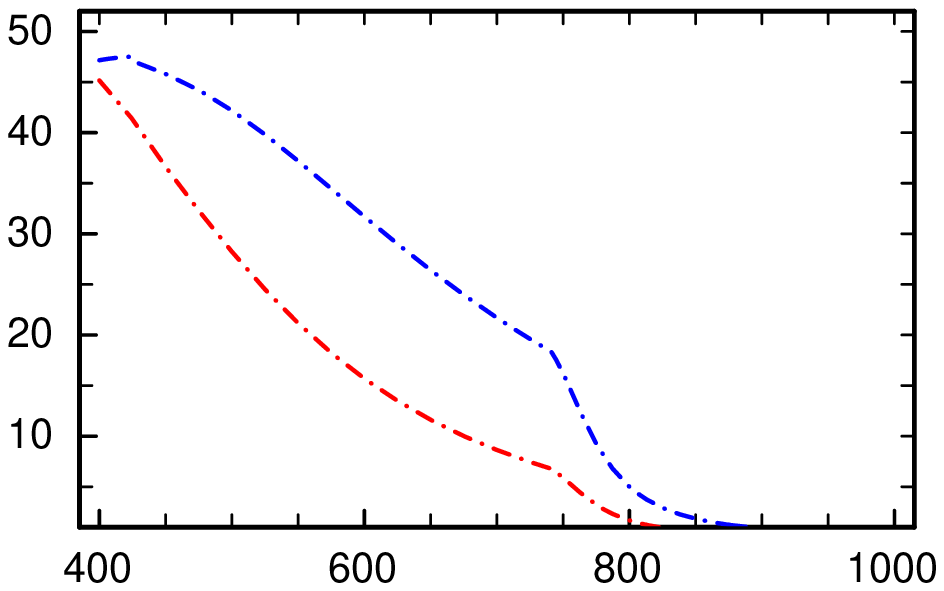,height=5cm}}}
\put(45.5,61){\mbox{$M_{\ti Q}~[{\rm GeV}]$}}
\put(45.5,0.5){\mbox{$M_{\ti Q}~[{\rm GeV}]$}}
\put(9,82){\makebox(0,0)[br]{{\rotatebox{90}{BR\,($\st_2$)~[\% ]}}}}
\put(9,22){\makebox(0,0)[br]{{\rotatebox{90}{BR\,($\sb_2$)~[\% ]}}}}
\put(44,101){\mbox{\scriptsize $\st_1 A^0$}}
\put(26,98){\mbox{\scriptsize $\st_1 Z^0$}}
\put(54,77.5){\mbox{\scriptsize $\st_1 h^0$}}
\put(30.5,81){\mbox{\scriptsize $\sb_1 W^+$}}
\put(43,40){\mbox{\scriptsize $\st_1 W^-$}}
\put(39.4,29){\mbox{\scriptsize $\st_1 H^-$}}
\end{picture}
\end{center}
\caption{ $M_{\ti Q}$ dependence of $\st_2$ and $\sb_2$ decay 
branching ratios for $M_{\ti U}=\frac{8}{9}\,M_{\ti Q}$, 
$M_{\ti D}=\frac{10}{9}\,M_{\ti Q}$, $A_t=A_b=400\gev$, 
$\mu=-1000\gev$, $M=300\gev$, $m_A=150\gev$, 
and $\tan\beta=3$. }
\label{fig:four}
\end{figure}


\end{document}